\RequirePackage{fix-cm}
\documentclass[smallextended]{svjour3}       
\usepackage{amsmath}
\usepackage[colorlinks=true]{hyperref}
\usepackage{amsfonts}
\usepackage{graphicx}
\newcommand{\citenum}[1]{\cite{#1}}
\makeatletter
\DeclareFontFamily{OMX}{MnSymbolE}{}
\DeclareSymbolFont{MnLargeSymbols}{OMX}{MnSymbolE}{m}{n}
\SetSymbolFont{MnLargeSymbols}{bold}{OMX}{MnSymbolE}{b}{n}
\DeclareFontShape{OMX}{MnSymbolE}{m}{n}{
    <-6>  MnSymbolE5
   <6-7>  MnSymbolE6
   <7-8>  MnSymbolE7
   <8-9>  MnSymbolE8
   <9-10> MnSymbolE9
  <10-12> MnSymbolE10
  <12->   MnSymbolE12
}{}
\DeclareFontShape{OMX}{MnSymbolE}{b}{n}{
    <-6>  MnSymbolE-Bold5
   <6-7>  MnSymbolE-Bold6
   <7-8>  MnSymbolE-Bold7
   <8-9>  MnSymbolE-Bold8
   <9-10> MnSymbolE-Bold9
  <10-12> MnSymbolE-Bold10
  <12->   MnSymbolE-Bold12
}{}

\let\llangle\@undefined
\let\rrangle\@undefined
\DeclareMathDelimiter{\llangle}{\mathopen}%
                     {MnLargeSymbols}{'164}{MnLargeSymbols}{'164}
\DeclareMathDelimiter{\rrangle}{\mathclose}%
                     {MnLargeSymbols}{'171}{MnLargeSymbols}{'171}
\makeatother

\newcommand{\inner}[2]{\left\llangle #1 \vphantom{#2} ,
  #2 \vphantom{#1} \right\rrangle} 
 
\newcommand*{\pd}[2]{\frac{\partial{#1}}{\partial{#2}}}
\newcommand*{\pdel}[2]{\frac{\delta{#1}}{\delta{#2}}}

\newcommand*{\tr}[1]{\operatorname{tr}\left[ #1 \right]}
\newcommand*{\avg}[1]{\left\langle #1 \right\rangle}

\begin{document}

\title{ Dual Characterization of the Ornstein-Zernike Equation in Moment Space}


\author{ David M. Rogers}
\institute{D. M. Rogers \at
University of South Florida, 4202 E. Fowler Ave., CHE 205, Tampa, FL 33620\\
Tel.: +1-813-9744298\\
\email{davidrogers@usf.edu}}

\date{Submitted April 7, 2018.}

\maketitle

\begin{abstract}
  The molecular density functional theory of fluids provides an exact
theory for computing solvation free energies in implicit solvents.
One of the reasons it has not received nearly as much
attention as quantum density functional theory for implicit electron densities
is the paucity of basis set expansions for this theory.
This work constructs a minimal Hilbert space version of the Ornstein-Zernike
theory over the complete spatial, rotational, and internal conformational space
that leaves the choice of basis open.
The basis is minimal in the sense that it is isomorphic to a choice of
molecular property space (i.e. moments of the distribution), and does
not require auxiliary grids.
This can be exploited, since there are usually only a few `important'
properties for determining the structure and energetics of a molecular fluid.
Two novel basis expansions are provided which emphasize either
the multipolar expansion (most useful for flexible molecules)
or the rotational distribution (most useful for rigid bodies described using quaternions).
The perspective of this work shows that radial truncation of
the Bessel series over translation space
determines an analytical extrapolation of these
functions to the origin in reciprocal space.
We provide a new density functional theory that naturally fits the moment-based,
matrix approach.
Three diverse applications are presented: relating the present approach
to traditional rotational invariants, demonstrating the stability of convex optimization
on the density functional, and finding analytical expression for dispersion
contributions to the solvation free energies of point polarizable dipoles.
\end{abstract}

\section{ Introduction}

  In this work, we show that a full expansion in terms of local,
molecular fields is `dual' to a basis expansion of the density function.
The duality between these two approaches is exactly that between a distribution and its
moments, and is a special case of the moment problem in classical probability theory.\cite{hdett97}
Hence, the rotational invariant expansion of Blum\cite{lblum72}
is only a particular choice which is best used to describe multipole moments
of a rigid molecule.
We provide two examples of moment spaces which use this
transformation to eliminate rotational density space in favor of observable
molecular properties.

  The connection to moments of the density distribution was
was implicit in the original spherical harmonic expansion of Blum,\cite{lblum72}
but seems to have been first used effectively by Sluckin.\cite{tsluc81}
Nevertheless, the spherical invariant expansion has been the framework
adopted by most later works.  Interestingly, many of those later
works find convenient expressions in matrix form.\cite{lblum90}
When using general potentials, Fries and coworkers noted there were difficulties
associated with carrying out six-dimensional angular
integrals (over rigid-body rotational degrees of freedom).\cite{pfrie86}

  More recent works have shown the usefulness of expressing
the translational part of the Ornstein-Zernike (OZ)
equation in a Cartesian grid basis.\cite{mikeg95,akova00,rishi13}
Most of those works, however, apply to the context
of RISM theory.  That theory has advanced
to a great degree,\cite{svolo12,jjohn16} but
requires describing each atom, rather
than each molecule, as a separate site.

  Work by Borgis and collaborators\cite{rrami02,lding17}
has shown the usefulness of coarsening the rotational distribution
into a spatial dipolar polarization density.
Those works outline a path between the spherical invariant
and spatial number and dipolar density representations that makes it
possible to phrase solvation thermodynamics
entirely in terms of latter, at least for the
Stockmayer fluid.  Its application to liquid water has
turned out to be very successful,\cite{szhao13} motivating extension
to higher orders.\cite{gjean16}

  The mathematical motivation for this re-phrasing is the relative difficulty of working with
probability distributions over molecular position and conformation space,
$x = (r,q,s) \in \Omega_1 \equiv \mathbb R^3 \times \text{SO}(3) \times \mathbb S$,
where $r$ is the center of mass location, $q$ is a unit quaternion describing the orientation,
and $s \in \mathbb S$ describes all internal coordinates.
The set $\mathbb S$ is an arbitrary space of internal molecular coordinates.
In applications, the space of distribution functions over $\Omega_1$
is usually truncated to a finite number of basis functions.
Thus, it is natural to look for physically motivated
transformations of those basis functions that give more insight into
molecular structure.

  By following this idea through, we show how existing literature on one-component
fluids of spherical particles can, in most cases, be lifted to a matrix notation
valid for molecules with internal and rotational degrees of freedom.
This opens new ideas for analytical and numerical investigations.
Important potential application areas include examining the density-dependence
of dielectric effects,\cite{dmaty94,frain01,kdyer08}
simplifying models of solvation\cite{dbegl97,nrft,lvrbk09}
including Casimir (density fluctuation) forces,\cite{pbuen05}
and extending dynamic density functional theories.\cite{aarch09}

  This work is divided into sections presenting I,II) the linear transformations
between traditional projections of the 1-particle density onto a basis and
field equations for average molecular properties, and III,V) a formulation
of OZ and classical fluid density functional theory (DFT)
directly in molecular field space.
Section IV presents two alternative basis expansions -- multipolar moments
appropriate for general molecules
and a quadrature basis for distributions over quaternions appropriate for rigid bodies.
Applications are presented in Sec. VI, where
we recover in a few steps the mean-spherical approximation
for dipolar hard spheres and then show how it can be adapted to
investigate Casimir forces in fluids of polarizable point dipoles.


\section{ Problem Statement}


  The usual perturbation theory is most easily stated by defining
functionals,
\begin{equation}
Z[\Phi] \equiv \avg{ e^{\inner{\Phi}{\hat\rho}} }, \quad
Z_N[\Phi] \equiv \avg{ e^{\inner{\Phi}{\hat\rho}} | n = N }
,\label{e:Z}
\end{equation}
where $\Phi(x)$ is a (real-valued) function of the coordinates, $x$, of one molecule
and $\hat\rho(x)$ is the instantaneous number density of molecules
at point $x$ (a function of the microstate).
Single-brackets denote an average over the Grand-Canonical ensemble,
and double-brackets indicate the inner product,
\begin{equation}
\inner{\Phi}{\hat\rho} \equiv \int_{\Omega_1} dx \; \Phi(x) \hat\rho(x) .
\label{e:inner}
\end{equation}

  Then, since Eq.~\ref{e:Z} is a moment generating function for $\hat\rho$,
the 1-particle density is its first cumulant,
\begin{equation}
\rho(x | \Phi) = \pdel{\log Z[\Phi]}{\Phi(x)}
.\label{e:rho}
\end{equation}
Thermodynamic integration then provides a method for calculating
the density response to `charging' $\Phi_\lambda$
as $\lambda$ smoothly changes $\Phi_0 = 0$ into $\Phi_1 = \Phi$,
\begin{equation}
\rho(x | \Phi) = \rho(x | 0) + \int_0^1 d\lambda\;
\inner{\pd{\Phi_\lambda}{\lambda}}{\pdel{\rho(x|\Phi_\lambda)}{\Phi_\lambda}}
.\label{e:drho}
\end{equation}

  It is usually the case that the pair potential can be decomposed into relative
translation and internal coordinate-dependent parts,
$\Phi(x) \asymp \sum_{i=1}^M \phi_i(r) f_i(q,s)$, where $f : \text{SO}(3)\times \mathbb S \to \mathbb R^M$
is some vector-valued function of the molecular conformation,
$\phi : \mathbb R^3 \to \mathbb R^M$ is a vector-valued potential,
and $\asymp$ denotes convergence with increasing expansion order, $M$.
In this case, we can work out conditions where
\begin{equation}
\inner{\Phi}{\hat \rho} = \inner{\phi}{\hat F}, \label{e:equiv}
\end{equation}
which explicitly uses the field density of molecular observables, $f$,
\begin{equation}
\hat F(r) \equiv \iint dq ds f(q,s) \hat\rho(r,q,s)
 \asymp \sum_{j=1}^n \delta(r,r_j) f(q_j,s_j)  \label{e:F}
\end{equation}
The new inner product uses the obvious definition,
\begin{equation}
\inner{\phi}{\hat F} \equiv \int dr \phi^\dagger(r) \hat F(r) \asymp \sum_{j=1}^n \phi^\dagger(r_j) f(q_j,s_j)
,\label{e:dE2}
\end{equation}
where $z^\dagger$ denotes the complex-conjugate transpose of $z$.
Because of this equivalence, we abuse notation to write $Z[\phi] = Z[\Phi]$.

  It is equally valid to write thermodynamic integration in terms of the
``molecular field,'' $F(r|\phi) = \delta \log Z[\phi] / \delta \phi(r)$,
\begin{equation}
F(r | \phi) = F(r | 0) + \int_0^1d\lambda\;
\inner{\pd{\phi_\lambda}{\lambda}}{\pdel{F(r|\phi_\lambda)}{\phi_\lambda}}
.\label{e:dF}
\end{equation}
Providing a useful transformation between the first and second versions of the perturbation theory
giving the integrand of Eq.~\ref{e:dF} is the motivation for this work.
These relations are then used to define a density-operator
algebra and to re-state the OZ and DFT theories in moment space.
We will find that the new algebra provides a natural setting
for posing convex optimization problems that arise in solving
typical problems in fluid density functional theory.

\section{ Transformation Between Formalisms and Convolution Algebra}

  The two formalisms are of course related by the requirement
that the field be the average of the distribution,
$F(r) = \iint dq'ds'\; f(q',s') \rho(r,q',s')$.
This is a direct consequence of the definition in Eq.~\ref{e:F}.
Thus, a perturbation theory for $\rho$ can be used to find one for $F$.

  We prove in this section that the correspondence is actually one to one
whenever a finite basis is used to represent the distribution over
conformation space as a function in the Hilbert space, $\mathcal H^\text{conf}$.
This finite basis then determines the property vector, $f$, up to linear transformation.
Obviously, this requires that the density basis for $\mathcal H^\text{conf}$ have dimension, $M$,
equal to that of $f$.
The key relation is given by,
\begin{equation}
\iint dq ds\; \Phi(x) \delta(x, x') = \delta(r, r') \sum_{i=1}^M \bar \phi_i(r) f_i(q',s') 
.\label{e:phi}
\end{equation}
The overbar denotes complex conjugation,
although we almost always work with real $\phi,f$.
Note that Eq.~\ref{e:phi} implies Eq.~\ref{e:equiv}, which proves both
operators, $\Delta \hat E$, and the corresponding free energy functionals, $Z$,
are equivalent.

  To prove our assertion above, we must show the validity of Eq.~\ref{e:phi}
and that the translation from $\rho$ to $F$ (Eq.~\ref{e:F})
can be inverted to find $\rho$ from $F$ again.
Since an $M$-dimensional basis for density functions, $\rho$, implies
a choice for the reproducing kernel,
\begin{equation}
\hat\rho(x) = \int dx' \delta(x,x') \hat\rho(x')
,\label{e:rrho}
\end{equation}
we can make use of the fact that the span of $\delta(q,s; q',s')$
traces out the $M$-dimensional Hilbert space, $\mathcal H^\text{conf}$,
and write its orthonormal basis as $\{Y_i\}_1^M$.\cite{naron50}
Eq.~\ref{e:phi} is then proved directly by writing the reproducing kernel
as the resolution of identity,
\begin{equation}
  \delta(q,s; q',s') = \sum_{i=1}^M \bar Y_i(q,s) Y_i(q',s') =: u^\dagger(q,s) f(q',s')
.\label{e:eigen}
\end{equation}
Substituting Eq.~\ref{e:eigen} into Eq.~\ref{e:phi} and comparing
to $\phi^\dagger f$ gives the identifications,
\begin{align}
f(q,s) &= A Y(q,s) \label{e:fn} \\
u(q,s) &= A^{\dagger -1} Y(q,s) \label{e:un} \\
\phi(r) &= \iint dq ds\; u(q,s) \Phi(x) \label{e:pn}
,
\end{align}
for any invertible, $M\times M$ matrix, $A$.

  To find $\rho$ from $F$, insert Eq.~\ref{e:eigen} into
Eq.~\ref{e:rrho} to find,
\begin{equation}
\hat \rho(x) = \int dx' \; \delta(r,r') u^\dagger(q,s) f(q',s') \hat\rho(x')
  = u^\dagger(q,s) \hat F(r). \label{e:form}
\end{equation}
The last step follows from Eq.~\ref{e:F} and the fact that $\hat F_i(r) \in \mathcal H^\text{trs}$,
the Hilbert space of functions with reproducing kernel, $\delta(r,r')$.

  The above considerations show formally how OZ theory on density
space corresponds exactly to a ``molecular OZ'' theory on moment space
because both are built on the same perturbation expansion of $\log Z$,
and an invertible linear transformation to moment space can be constructed
for any space of density functions which admits a convergent basis expansion.
This transformation provides a dual characterization of the moment space.
Note that the distributions $\rho(x)$ are nonnegative functions.
Further, for each $r$, they belong to some proper cone, $\mathcal C(r)$,
generated by the conformations, $(q,s)$, reachable by molecules located
at $r$.  The moment space at $r$ must satisfy $u^\dagger(q,s) F(r) \in \mathcal C(r)$.

\subsection{ Convolution Algebra}\label{s:conv}

  The change of basis equations, Eqns.~\ref{e:fn}-\ref{e:un}
show that the space of moment functions, $F(r)$, is isomorphic to
the set of density functions, $\rho(x)$ restricted to a particular
$\mathcal H^\text{trs}\otimes\mathcal H^\text{conf}$.
This section develops the corresponding expressions
for second moments.

  In this representation, convolution is represented as matrix multiplication
in the operator algebra generated by,\footnote{The `*' denoting convolution is
omitted when two matrices appear next to one another.  This should
not cause confusion since matrices are written with uppercase letters
and we never use pointwise products between them.}
\begin{equation}
(C R)(r,r') \equiv \int dr'' \; C(r,r'') R(r'',r')
,\label{e:CH}
\end{equation}
which is isomorphic to ordinary convolution,
\begin{equation}
(c*\rho_2)(x,x') \equiv \int dx''\; c(x,x'') \rho_2(x'',x')
,
\end{equation}
when $c$ is an energy-type function and $\rho_2$ is a density-type function,\footnote{  Note how the result of Eq.~\ref{e:CH} is of `mixed type.'
In the reproducing kernel Hilbert space context, we could cheat and
insert the change of basis $f(q,s) = (AA^{\dagger}) u(q,s)$.
However, the difference in types will likely be important
for convergence of inner-products between energy-type and density-type functions.}
\begin{align}
c(x,x') &\asymp f(q,s)^\dagger C f(q',s') \iff C = \iint dq ds \iint dq'ds'\; u(q,s) u^\dagger(q',s') c(x,x') \label{e:ent} \\
\rho_2(x,x') &\asymp u(q,s)^\dagger R u(q',s') \iff R = \iint dq ds \iint dq' ds'\; f(q,s) f^\dagger(q',s') \rho_2(x,x')
.\label{e:denst}
\end{align}

  The inner product can be extended to operators
by defining a star operation which takes the
conjugate-transpose of each matrix and interchanges source
and destination points ({\em i.e.} $R_{ij}^*(r,r') = \bar R_{ji}(r',r)$
and $\rho_2^*(x,x') = \bar \rho_2(x',x)$), and a trace operation,
\begin{align}
\tr{C} &\equiv \frac{1}{V} \int dr \; \sum_{i=1}^M C_{ii}(r,r),\quad
\tr{c} = \frac{1}{V} \int dx \; c(x,x) \label{e:tr} \\
\intertext{so that}
\inner{C}{R} &= \tr{ C^* R } = \tr{c^* * \rho_2}
. \label{e:opinner}
\end{align}
Symmetry on interchanging source and destination points
makes all pairwise operators selfadjoint, satisfying $C = C^*$.
Treating multi-component solutions by adding
an index for each molecule type
(e.g. $C_{\alpha i,\gamma j}$) is a trivial extension.

  These extensions are helpful for formulating
a complete theory of pair correlations.
Because of the division by $\rho$, the usual total correlation function,
\begin{equation}
h(x,x') \equiv \frac{\avg{\sum_{\stackrel{i,j = 1}{i\ne j}}^{n} \delta(x, x_i) \delta(x', x_j)}}{\rho(x)\rho(x')} - 1. \label{e:h}
\end{equation}
is not easily used with $\mathcal H^\text{trs}\otimes\mathcal H^\text{conf}$.
Instead, we define the matrix-valued total correlation function, $H$,
through
\begin{align}
(R H R)(r,r') &\equiv \iint dq ds \iint dq'ds'\; \rho(x) h(x,x') \rho(x') f(q,s) f^\dagger(q',s')
,\label{e:rhr}
\intertext{with}
R(r,r') &\equiv \iint dq ds \iint dq'ds'\; \rho_2(x,x') f(q,s) f^\dagger(q',s') \\
&\asymp n(r) \delta(r - r') \avg{f_j(q,s) f_j^\dagger(q',s') | r_j = r}, \text{ where} \\
\rho_2(x,x') &\equiv \avg{ \sum_{j=1}^n \delta(x,x_j) \delta(x',x_j) }
= \avg{\Delta\hat\rho(x)\Delta\hat\rho(x')} - \rho(x)h(x,x')\rho(x')
\end{align}
is a 1-particle correlation function.
This allows working completely in terms of our Hilbert space.

  In the special case of pairwise energy functions of the form,
\begin{equation}
u_2(x,x') = f(q,s)^\dagger {U_2}(r,r') f(q',s')
,
\end{equation}
The average pair interaction energy converges to
\begin{equation}
U_\text{pair} = \frac{1}{2} \avg{\sum_{\stackrel{i,j = 1}{i\ne j}}^{n} u_2(x_i,x_j)}
   \asymp \frac{V}{2} \inner{U_2}{(R H R)(r,r') + F(r) F^\dagger(r')}
,\label{e:U2}
\end{equation}
as the size of $\mathcal H^\text{trs}$ increases.


  To relate Eq.~\ref{e:drho} to Eq.~\ref{e:dF},
we define the field-field correlation function as,
\begin{equation}
Q(r,r') = \pdel{F(r)}{\phi(r')} = \avg{\Delta \hat F(r) \Delta \hat F^\dagger(r')}
= (R H R)(r,r') + R(r,r') \label{e:Q}
.
\end{equation}
In our notation, the rows of $Q$ correspond to the numerator and the columns to the denominator.
These are density-type functions, so Eq.~\ref{e:denst} applies
to transform between $Q,RHR,R$
and $\avg{ \Delta \hat \rho(X) \Delta \hat \rho(X') | \Phi},\rho^2 h,\rho_2$.
Using this transformation in Eq.~\ref{e:dF} verifies that the response of the field vector
is the average of the density response to the corresponding potential
and {\em vice-versa} that the density response to a perturbation can be expanded
in a basis of functions whose coefficients depend linearly on the field response at each point.

  It should also be noted that $Q(r,r')$ is related to the
structure factor probed by x-ray, electron and neutron
scattering experiments.\cite{davis,gmaze06}
Because each of those experiments probes a single
type of density (electric or atomic), each provides one
linear constraint on the Fourier transform, $\iint e^{-ik\cdot (r-r')} Q(r,r') dr dr'$
for each measured wavevector, $k$.


\section{ Useful Basis Expansions}\label{s:basis}

  Any of the relations in Eq.~\ref{e:eigen}, \ref{e:fn}, or~\ref{e:pn} could
be used to derive a choice for the basis expansion to be used in a numerical
or theoretical perturbation theory.  However, it is usually unnecessary
to start from the eigendecomposition of identity in Eq.~\ref{e:eigen}.
Instead, most kernels are derived physically by decomposing the interaction
energy expression using Eq.~\ref{e:phi}.  This leads to a set of `important'
moments, $f$, which are the first few terms of an expansion of the interaction
energy.  Next, these moments are used to find appropriate kernels,
$\delta$, by writing $\rho$ in the form of Eq.~\ref{e:form}.

  The obvious basis for a dipole is,
\begin{equation}
  f^\text{dip}(q,s) = \begin{bmatrix}
  1 \\ q \mu(s) \bar q
  \end{bmatrix},\label{e:fdip}
\end{equation}
where $q \in SO(3)$ is treated as a quaternion to rotate the molecule's dipole
moment, $\mu$, expressed as a function of the molecule in
a reference orientation with internal coordinates, $s$.
In this basis, it is possible to express both electrostatic and isotropic
interactions up to dipolar order.  It has been investigated extensively
for rigid molecules (where $\mu$ is a fixed axis).\cite{gjean16}

  We show here that a systematic approach for other situations
can be based on finding a quadrature rule for Eq.~\ref{e:eigen},
where $\delta$ is the reproducing kernel for a class of functions
on `important' or `coarse-grained' sub-spaces of $\text{SO}(3)\times \mathbb S$.
In that case, all integrals over the coarse-grained
space including Eq.~\ref{e:phi} and Eq.~\ref{e:F} can be
written as sums over quadrature points, $\xi_i \in \text{SO}(3)\times\mathbb S$,
and we only ever need to know the values of the densities, $\rho$,
and interaction energies, $\Phi$, at those points.

  Specifically, assume we are given an $M$-dimensional Hilbert space of functions on
$\text{SO}(3)\times \mathbb S$ with reproducing kernel $\delta(q,s; q',s')$
and an $M$-point quadrature rule for the space.
That quadrature rule should have weights $w_i$ at points
$\xi_i = (q_i,s_i)$ for which,
\begin{equation}
\delta(q,s; q',s') = \sum_i w_i \delta(q,s; \xi_i) \delta(\xi_i; q',s')
.
\end{equation}
In this case, choose
\begin{equation}
f_i^\text{rk}(q,s) = w_i \delta(q,s; \xi_i) \;\text{ and }\; u_i^\text{rk}(q,s) = \delta(q,s; \xi_i)
,\label{e:frk}
\end{equation}
so Eq.~\ref{e:eigen} is automatically satisfied.
The next two equations (\ref{e:pn} and~\ref{e:fn})
identify $\phi_n(r)$ and $f_n$ as weighted sums of the external potential
$\Phi(r,\xi_i)$ and the density operator, $\hat\rho(r,\xi_i)$, respectively,
evaluated at those quadrature points.

  This idea leads to a physically useful basis for full multipolar moment expansion
up to order $T = \sqrt{M}-1$ using the equivalent point charge representation,\cite{droge15}
\begin{equation}
f_i^\text{pc}(q,s) = w_i \int_{y \in \mathbb R^3} d\rho^\text{chg}(y; s)\; K(\xi_i, q y \bar q)
,\label{e:fpc}
\end{equation}
where $\xi_i \in S^2$ are Lebedev or other useful\cite{cahre09}
quadrature points on the unit sphere, $S^2$, for polynomials up to order $T$ and
$K$ is the reproducing kernel defined in Ref.~\citenum{droge15}.
Equation~\ref{e:fpc} provides an equivalent point charge, $f_i^\text{pc}$, at point $\xi_i$,
in the laboratory frame from a molecule in a reference frame 
with charge distribution, $d\rho^\text{chg}(y; s)$.

  An example illustrates the idea.  For $T=1$, the reproducing kernel is $K(x, y) = (1 + 3 x\cdot y)$.
It can be integrated with four quadrature points (of equal weight)
at the vertices of a tetrahedron inscribed in a unit sphere.
These are related to the choice of basis in Eq.~\ref{e:fdip}
by the matrix,
\begin{equation}
f^\text{pc,1}(q,s) = \frac{1}{4} \begin{bmatrix}
1 & 0 & 0 & 3 \\
1 & \sqrt{8} & 0 & -1 \\
1 & -\sqrt{2} & \sqrt{6} & -1 \\
1 & -\sqrt{2} & -\sqrt{6} & -1 
\end{bmatrix} f^\text{dip}(q,s).
\end{equation}
The resulting moments describe
the effective point charge placed at each vertex.
The first column distributes the total charge evenly
among the four points.  The last three are row-vectors of the vertex locations.
It can be checked that they sum to zero
since a dipole should not contribute to the total charge.

  As another example, for rigid molecules, it may make sense to choose instead
\begin{equation}
f_i^\text{rot}(q,s) = \tfrac{1}{12} K_R(q_i, q)
,
\end{equation}
where $q_i$ are the 12 Hurwitz quaternions\cite{skony98} (modulo an overall sign, which
express two rotations to each of the vertices of an octahedron) and
\begin{equation}
K_R(q, q') = \sum_{j=0}^{5} (j+1) U_j(q \cdot q')
,\label{e:KR}
\end{equation}
is the reproducing kernel for the set of polynomial functions on
the unit quaternions with degree less than $6$.
The scalar product between quaternions is,
\begin{equation}
q\cdot q' \equiv (q\bar q' + q' \bar q) / 2, \label{e:dot}
\end{equation}
and $U_j$ is the $j^\text{th}$ order Chebyshev polynomial of
the second kind,
\begin{equation}
U_0(z) = 1, \quad U_1(z) = 2z, \quad U_j(z) = 2z U_{j-1}(z) - U_{j-2}(z)
.
\end{equation}

  The reproducing property of Eq.~\ref{e:KR} is a consequence
of the zonal expansion\cite{sboch54} and the Funk-Hecke formula.
For working with Eq.~\ref{e:dot}, it is helpful to have the geometric
identities,
\begin{equation}
a\cdot b = (ac)\cdot (bc) = (ca)\cdot (cb)
,
\end{equation}
for any normalized quaternion $c$ (i.e. $c \bar c = 1$).
These express invariance of Eq.~\ref{e:KR} to the choice of
laboratory orientation frame.

  The dipole representation has only recently been investigated in the literature
in the case of rigid molecules.  Our generalizations to multipolar distributions with
flexible molecules and over the quaternion algebra have not been investigated
before for density functional theories.
These have the potential to offer order of magnitude computational savings
over straightforward tensor products of angular grids.\cite{lding17}

\subsection{ Translational Basis}

  By writing most of our results as functions of $r,r'$ we have
been able to avoid the question of a translational basis.  However,
extension of the matrix formulation to $\mathcal H^\text{trs}$ is
indispensable for numerical work.

  The simplest translational basis to implement is the 3D Fourier basis
on a regularly spaced grid representing a periodic cube with lattice spacing $L$
(so $\Omega_1^\text{FT} = [0,L)^3 \times \text{SO}(3) \times \mathbb S$).
This case (for arbitrary 3D lattices) along with a fast and accurate numerical
approximation for low-wavevector modes
was discussed by Essman et. al.,\cite{uessm95}
and additional results are available in Ref.~\citenum{droge18}.
It corresponds to a reproducing kernel that is `band-limited'
by truncation at a cutoff $K_\text{max}$, so that the
reproducing kernel is,
\begin{equation}
\delta^\text{FT}(r,r') = \frac{1}{K_\text{max}^3}
\sum_{k_1,k_2,k_3=1}^{K_\text{max}} e^{2\pi i k\cdot (r - r') / L}
.
\end{equation}
Note that this is a smooth analytic function of $r$ and $r'$, but
its use in numerical evaluations of $-\log Z[\phi]$ implies the projection of
pair energy functions via $\phi(r) = -\beta \int \delta(r,r') U(r') dr'$.
Our argument shows that the perturbation expansion, when written
in matrix form, is exact for such $\phi \in \mathcal H^\text{trs}$.

  The traditional basis for the OZ equations has been the spherical
Bessel function expansion.  We provide a new formula for the expansion here
that resolves the long-standing, open question on how to properly treat the origin.
Our answer is based on the simple analogy between the discrete Fourier transform
on the torus and the Fourier-Bessel series for functions on finite intervals.
The spherical Bessel function method is usually supplemented by the
condition that the functions concerned (e.g. $ h(r)$) have finite range $R_\text{max}$,
so $\Omega_1^\text{SB} = ([0,R_\text{max}]\times S^2) \times \text{SO}(3) \times \mathbb S$.
Adding the condition that $h(r > R_\text{max}) = 0$ with $h$ continuous at $R_\text{max}$
is exactly the prerequisite for applying the Fourier-Bessel series.


  Truncating the series at $N_\text{max}$ and using a spherical harmonic
expansion over directions, $\vec r \in S^2$ leads to the general expression,
\begin{equation}
h(r) = \sum_{n=1}^{N_\text{max}} \sum_{l=0}^{L_\text{max}} \sum_{m=-l}^l C_{n,l,m} h_{n,l,m}(r),
\quad h_{n,l,m}(r) \equiv
j_l(|r| z_{n,l}/R_\text{max}) Y_l^m(\vec r)
,\label{e:SB}
\end{equation}
where $z_{n,l}$ is the $n^\text{th}$ smallest root of $j_l(z)$ (not counting roots at $z=0$).
The coefficients are the real values of $h_r$.
Our normalization of the spherical harmonics is
$\int_{S^2} d\vec r \, \bar Y_{l'}^{m'}(\vec r) Y_l^m(\vec r) = 4\pi \delta_{l,l'}\delta_{m,m'}$.

  Functions in this Hilbert space have the Fourier transform,
\begin{align}
\tilde h_{n,l,m}(k) &= 4\pi R^3 i^{-l} Y_l^m(\vec k)
\int_0^1 x^2 dx j_l(|k| R x) j_l(x z_{n,l}) \notag \\
&= 4\pi R^3 i^{-l} Y_l^m(\vec k)
\frac{- z_{n,l} j_{l-1}(z_{n,l}) j_l(|k|R)}{z_{n,l}^2 - (|k|R)^2}
.\label{e:FSBT}
\end{align}
On shells where $R |k| = z_{n',l}$, this reduces to
zero if $n \ne n'$ or (for $n = n'$)
\begin{equation}
\tilde h_{n,l,m}(\vec k z_{n,l}/R) = 2\pi R^3 i^{-l} Y_l^m(\vec k)
\left( - j_{l-1}(z_{n,l}) j_{l+1}(z_{n,l}) \right)
.\label{e:SBT}
\end{equation}
Each of the $h_{n,l,m}$ are mutually orthogonal, with normalization
\begin{equation}
N_{n,l,m} \equiv \int dr \bar h_{n,l,m}(r) h_{n',l',m'}(r) = \delta_{n,n'}\delta_{l,l'}\delta_{m,m'}
2\pi R^3 j_{l-1}(z_{n,l})^2
.
\end{equation}
The $l=0$ case makes use of $j_{-1}(z) = \cos(z)/z$.
The associated reproducing kernel is therefore
\begin{equation}
\delta^\text{SB}(r,r') = \sum_{n=1}^{N_\text{max}} \sum_{l=0}^{L_\text{max}} \sum_{m=-l}^l
\frac{h_{n,l,m}(r) \bar h_{n,l,m}(r')}{N_{n,l,m}}
.
\end{equation}

  Note that although the spherical Bessel function basis makes physical sense
for smooth radial functions with limited range,
it is not closed under convolution or multiplication of functions.
Nevertheless, Eq.~\ref{e:SBT} shows that the coefficients
can be visualized as momentum shells, and so convolution
of $h$ and $h'$ can be projected back to the Hilbert space (Eq.~\ref{e:SB})
by computing the values of $\tilde h_{n,m,l} \tilde h'_{n',m',l'}$
on those shells using Eq.~\ref{e:FSBT}
(which has the same form as the `SB' reproducing kernel in Fourier space).

\section{ Transformation of OZ and DFT}\label{s:OZ}

  This section derives the well-known connection
between correlation functions and effective Green's functions for field theories.
The statement leads to both an OZ equation defined directly
in matrix space and a corresponding DFT.

  The Legendre-Fenchel transform pair for $\log Z[\phi]$ is,\cite{eshrig}
\begin{align}
\mathcal S[F] &\equiv \inf_\phi \left[ \log Z[\phi] - \inner{\phi}{F} \right] \label{e:S} \\
-\log Z[\phi] &= \inf_F \left[ -\mathcal S[F] - \inner{\phi}{F} \right] \label{e:DFT}
\end{align}
The notation $\mathcal S$ emphasizes that density functionals are generalized
entropies.  Because of the exponential function in Eq.~\ref{e:Z},
$\log Z$ is convex, infinitely differentiable (where $\log Z < \infty$)
and lower-semicontinuous, while $\mathcal S$
is concave and upper-semicontinuous.
The function space defined by all $\Phi(x)$ for which $\log Z[\Phi] < \infty$
is an Orlicz space.\cite{rao,wmaje14}
Since these are known to be convex, the second equality always holds
and we can state further that $|\inner{\phi}{F}| < \infty$
for functions satisfying $\log Z[\phi] < \infty$ and $S[F] > -\infty$.
In this paper, the spaces of $\phi$ and $F$ are both restricted to
$\mathcal H^\text{trs}\otimes \mathcal H^\text{conf}$.


\subsection{ OZ Equation}

  Eq.~\ref{e:Q} is related to the second derivative of $\mathcal S$ by the thermodynamic
conjugate relationship,
\begin{equation}
\pdel{\phi(r)}{F(r')} = Q^{-1}(r,r') = -\pdel{^2 \mathcal S[F]}{F(r) \delta F(r')}
.\label{e:iQ}
\end{equation}
The inverse indicated above is the inverse under convolution (Sec.~\ref{s:conv}).
In a field theory context, $Q^{-1}$ is the effective Green's function,
expressing the energy of interaction between densities at points $F(r)$
and $F(r')$ in a Gaussian approximation $\log P[F] \approx \mathcal S[F]$.

However, $\delta(r,r')$ is not necessarily diagonal in translation
or conformation space.
Hence, $\delta(x,x') u(x') \ne u(x)\delta(x,x')$ in general,
and we can only write $\rho_2(x,x') \asymp \delta(x,x') \rho(x)$
in the limit of an infinite basis size.


  To avoid assuming the limit, use the definitions in Eq.~\ref{e:Q} to find that
the equation $Q^{-1} Q = I\delta$ is solved by a pair of operators,
$R^{-1}$ and $C$, that satisfy,
\begin{equation}
H - C = H R C
,\label{e:MOZ}
\end{equation}
so that
\begin{equation}
Q^{-1} = R^{-1} - C
.\label{e:iQ2}
\end{equation}
Note that $Q$, $H$, and $R$ are all operators on
$\mathcal H^\text{trs}\otimes\mathcal H^\text{conf}$.
Thus, their inverses are formally defined while their existence depends on
whether the 1-particle density and conformational correlation,
$R$, is invertible. 

  Comparing to the OZ equation shows finally that
since $R, Q$ were defined as moments
(as in Eq.~\ref{e:fn}), then $R^{-1}$ and $C$ are
energy-type functions, whose translation to $\rho_2^{-1}$ and $c$
follows Eq.~\ref{e:ent}.
We note again that the convolution, $RHR$, rather than direct multiplication, is required
to maintain consistency with the use of the total correlation function, $h$, in the OZ equation
because $\delta(r,r') f(r') \ne f(r)\delta(r,r')$ for finite $\mathcal H^\text{trs}$.

\subsection{ Density Functional}

  Minimizing over $F$ in Eq.~\ref{e:DFT} for the fixed
potential field, $\Phi_{\gamma,y} = -\beta u_{2,\gamma}(x, y)$
of a molecule of type $\gamma$,
provides its excess chemical potential,\cite{bwido82,pdt2}
\begin{equation}
\beta\mu^\text{ex}_\alpha = -\int \rho_\alpha^0(y) dy \log Z[\phi_{\gamma,y}],
\label{e:muex}
\end{equation}
where $\phi_{\gamma,y}(r) = -\beta U_2(r, r_y) f(q_y,s_y)$.
Carrying out this minimization
will always result in a unique value of $\mu^\text{ex}_\alpha$
(along with its derivatives) due to the convexity properties stated earlier.
However, convergence of the minimization procedure
is a well-known, unsolved issue in practical calculations.\cite{lding17}
Our characterization of the space of $F$ (Eq.~\ref{e:form})
usually leads to a set of linear inequalities like $n(r) > 0$ that translate easily
to general convex optimization methods with provable numerical
convergence.


%

  To provide a full characterization, we derive a traditional,
explicit form for $\mathcal S[F]$
by integrating Eq.~\ref{e:iQ} along the path
$F_\lambda = F_0 + \lambda (F - F_0)$ from $F_0 = 0$ to
$F = F[\phi]$ and making an assumption for
the path-dependence of $Q^{-1}[F_\lambda] = Q^{-1}_\lambda$.
The functional derived in this section therefore
assumes at least that $\mathcal S[F_\lambda]$ is an analytic function of $\lambda$ along
this path.  The assumption is valid when the path does not cross a phase transition.

  The integral from $F_0$ (where $\phi = 0$ and
the entropy of solvation is $\mathcal S[F_0] = 0$) is,
\begin{equation}
\mathcal S[F] = -\int_0^1 (1-\lambda) d\lambda \; \inner{F - F_0}{Q^{-1}_\lambda * (F - F_0)}
,
\end{equation}
which uses the inner product from Eq.~\ref{e:dE2}.
Inserting Eq.~\ref{e:iQ2} decomposes this into ideal and excess parts,
\begin{align}
\mathcal S[F] &= \mathcal S_\text{id}[F]
                      + \tfrac{1}{2} \inner{F-F_0}{\bar C * (F - F_0)} \label{e:Dfunc} \\
\bar C &\equiv 2\int_0^1 (1-\lambda) d\lambda\; C_\lambda
.
\end{align}
Subscripts indicate the value of $\lambda$
at which each variable is evaluated.


  As noted in Ref.~\citenum{rrami02}, it makes sense
to treat the ideal, single-molecule term specially. 
In particular, in the limit where the basis exactly describes
the density function, the single-molecule term must go to,
\begin{align}
\mathcal S_\text{id}[F] &\asymp \int dx \left( \rho_1(x) - \rho_0(x) - \rho_1(x) \log \frac{\rho_1(x)}{\rho_0(x)} \right)
.\label{e:s2rho}
\end{align}
This entropy is exactly known when the 1-particle density function, $\rho(x)$ is also known,
but is incompletely determined by only the spatial, 1-particle moments, $F(r)$.

  We provide an alternative route to deriving the 1-particle term
by computing the entropy of an ideal gas polarized under a constant
external field, $\phi'$, relative to $d\mu^{(0)}_1(q,s)$ -- the probability distribution over
single-molecule conformations in the starting state, $F^0$ (with density $n_0$).
The gas' partition function is given by,
\begin{equation}
\log Z_1(\phi') = \log \sum_{n=0}^\infty \frac{e^{-n_0} n_0^n}{n!}
\left(\int e^{\phi'^\dagger f(q,s)} d\mu^{(0)}_1(q,s)\right)^n
= n_0 (Z_0(\phi') - 1)
.\label{e:Z1}
\end{equation} 
The ideal part of the entropy given a known average field $F(r)$ is then,
\begin{equation}
\mathcal S_\text{id}[F] =
\int dr \inf_{\phi' \in\mathbb R^M} \left[ n_0(r)(Z_0(\phi')-1) - \phi'^\dagger F(r) \right] \label{e:Sid2}
.
\end{equation}
A more physically intuitive derivation of Eq.~\ref{e:Sid2} could also
be provided by assuming the conformational correlation function, $R_\lambda$,
is identical to the maximum entropy result at each value of $F_\lambda$.

  Although Eq.~\ref{e:Sid2} appears odd at first, it reduces to the usual expression
when considering that $f$ must always contain one linear combination
that is constant so $n(r)$ can be determined from $F(r)$.
For an example, see $f^\text{dip}$ of Eq.~\ref{e:fdip}.
Separating $\phi$ this way provides
$\phi^\dagger f = \phi_n + \phi^\dagger_e f$ and $Z_0(\phi) = e^{\phi_n} Z_e(\phi_e)$,
so that $S^\text{id}$ is minimized when $n = n_0 Z_0$ and $F_e = n \pd{\ln Z_e}{\phi_e}$.

\begin{table}\sidecaption
\begin{tabular}{llrr}
\hline\noalign{\smallskip}
(mnl) & $T$ & $\Phi_T(\vec r, \vec \mu,\vec z)$
         & \phantom{space} $J_T$ \\
\noalign{\smallskip}\hline\noalign{\smallskip}
(000) & I & 1 & $j_0$ \\
(101) & P & $\vec z \cdot \vec r$ & $-i j_1$ \\
(011) & $\bar{\mathrm{P}}$ & $-\vec\mu \cdot \vec r$ & $-i j_1$ \\
(110) & $\Delta$ & $\vec z \cdot \vec \mu$ &$ j_0$ \\
(112) & D & $-3\Phi_P\Phi_{\bar P} - \Phi_\Delta$ & $-j_2$ \\
\noalign{\smallskip}\hline
\end{tabular}
\caption{ Rotational Invariant Basis Functions.  The
inter-molecular distance is $r = |r| \vec r$, where
unit vectors are written in bold letters.
The molecule at the origin is directed along $\vec z$, and
the molecule at $r$ is oriented in the $\vec \mu$ direction.
The first three columns label Blum's
triple-expansion in spherical harmonics,
while the last two columns give the Fourier transform
according to Eq.~\ref{e:FT}. With these definitions,
Fourier transformation replaces $\vec r$ with $\vec k$ in $\Phi_T$.}\label{t:inv}
\end{table}

\section{ Applications to the Dipolar Fluid}\label{s:dip}

  We give several applications of the present formalism.
First, the matrix formalism is compared with the traditional
expansion of the OZ equation into rotational invariants
used to solve the mean spherical approximation.
Next, we show the stability of the applying a primal-dual
interior-point method to Eq.~\ref{e:DFT}.
Finally, we use the new conformational dependence of Eq.~\ref{e:DFT}
to derive the Casimir energy for solvation of point polarizable
dipoles inside a polarizable dipole fluid.  Both of the latter
result represent substantial improvements over current
state of the art methods.

\subsection{ Comparison to Rotational Invariants}\label{s:wert}

  The rotational invariant basis functions\cite{lblum72}
are given in Table~\ref{t:inv}.
Each row of the table also lists the angular integral part of
its 3D Fourier transform in the form,
\begin{equation}
\int_{S^2} e^{-ik\cdot r} d\vec r\; \Phi_T(\vec r, \vec \mu, \vec z)
= 4\pi \Phi_T(\vec k, \vec\mu,\vec z) J_T(|r||k|). \label{e:FT}
\end{equation}
The notation $\vec r$ denotes a unit vector on $S^2$,
the surface of a sphere in 3D.

  Applying the choice of $f^\text{dip}$ from Eq.~\ref{e:fdip}
to a rigid molecule with dipole moment magnitude $m$,
\begin{align}
h(x,x') &= \begin{bmatrix}
1 \\
m q \vec z \bar q
\end{bmatrix}^T \begin{bmatrix}
h_I, & \vec r^T h_P \\
- \vec r h_P, & 3 h_D \vec r \vec r^T + I (h_\Delta - h_D)
\end{bmatrix} \begin{bmatrix}
1 \\
m q' \vec z \bar q'
\end{bmatrix} \label{e:hdip} \\
C(r,r') &= \begin{bmatrix}
c_I, & \vec r^T c_P \\
- \vec r c_P, & 3 c_D \vec r \vec r^T + I (c_\Delta - c_D)
\end{bmatrix}
.\label{e:Cdip}
\end{align}
where $\vec r$ is a unit vector in the direction $r$, and $H(r,r') = H(r - r')$
and $C(r,r') = C(r - r')$ are translationally invariant.
Note that taking the integral of Eq.~\ref{e:denst} multiplies
the inner matrix of Eq.~\ref{e:hdip} by the diagonal
factor, $B = \text{diag}(1,m^2/3,m^2/3,m^2/3)$, on both sides.
This matches the self-correlation function in a uniform system ($R(r,r') = \tfrac{\avg{n}}{V} \delta(r,r') B$),
leaving $H$ as the inner matrix in Eq.~\ref{e:hdip}.
The complementary function to $f$ is $u = f B^{-1}$.

  If we define the Fourier-Bessel transforms according to Eq.~\ref{e:FT},
\begin{equation}
\tilde c_T(|k|) \equiv 4\pi \int_0^\infty |r|^2 d|r|\; J_T(|r||k|) c_T(|r|)
,
\end{equation}
then the Fourier transforms of Eqns.~\ref{e:hdip}-\ref{e:Cdip} are the same, replacing
$\vec r$ with $\vec k$, and each $h_T$ with $\tilde h_T$ (and $c_T$ with $\tilde c_T$).
There is an essential symmetry here that interchanging
source and destination molecules (the star operation of Eq.~\ref{e:opinner})
leads to the same interaction energy.
Because of this, the Fourier transform, $\tilde C$, is Hermitian.

  It can be noticed immediately that
the lower-right block decomposes into a component,
$\tilde c_L \equiv \tilde c_\Delta + 2 \tilde c_D$ in the direction
$\vec k\otimes \vec k$ and $\tilde c_T \equiv \tilde c_\Delta - \tilde c_D$
along its orthogonal complement, $I - \vec k\otimes \vec k$.
These two components are scaled versions of $c_+$ and $c_-$
in Wertheim's original solution of the MSA for the dipolar fluid.\cite{mwert71}
The matrix notation of Eq.~\ref{e:MOZ} has thus lead us directly
to the three decoupled OZ equations given in that work (when $c_p = 0$).
The particular choice of scale used there was,
\begin{align}
\tilde h_L &= 6K' \tilde h_+ \label{e:hp} \\
\tilde h_T &= -3K' \tilde h_- \label{e:hm} \\
K' &\equiv \int_{0^+}^\infty \frac{dx}{x} h_D(x)
\end{align}
That scale was chosen to make the coefficient of $h_D$ be $1/3K'$,
so that $h_+(0) = h_-(0) = -1$.
We can summarize Wertheim's process of finding these
boundary conditions by noting that he found a general
relation between spherical Bessel transforms of orders 2 and 0,
\begin{align}
-\int_0^\infty x^2 dx j_2(k x) f(x) &= \int_0^\infty x^2 dx j_0(k x) \underline{f}(x) 
= \text{SBT}_0[\underline{f}](k) = \text{SBT}_2[f](k) \label{e:rel} \\
\underline{f}(x) &\equiv f(x) - 3\int_x^\infty \frac{dy}{y} f(y) = \text{SBT}^{-1}_0 [\text{SBT}_2[f]](x) .
\label{e:sbt2}
\end{align}
Eq.~\ref{e:rel} can be verified by substituting for $\underline f$,
interchanging the integrals and integrating by parts.
The last parts of Eqns.~\ref{e:rel} and~\ref{e:sbt2} use SBT$_n$ as shorthand for the
spherical Bessel transform of order $n$ using $J_T$ from
Table~\ref{t:inv}.  Thus,
\begin{equation}
h_+ = \frac{2}{6K'} \text{SBT}^{-1}_0[\text{SBT}_2[h_D]] + \frac{1}{6K'} h_\Delta
,
\end{equation}
and $h_+(r) = -1$ for $r$ within the excluded volume region, since $h_D(r) = h_\Delta(r) = 0$ there.

  The corresponding MOZ equations (\ref{e:MOZ}) read
\begin{align}
\tilde h_+ - \tilde c_+ &= 2K'\rho m^2 \tilde h_+ \tilde c_+ \\
\tilde h_- - \tilde c_-   &= -K'\rho m^2 \tilde h_- \tilde c_-
.
\end{align}
They differ from Wertheim's result because here $c$ should scale as $\beta/4\pi \epsilon_0 r^3$,
rather than $\beta m^2/4\pi \epsilon_0 r^3$.

\subsection{ Stability of Convex Optimization}

  In this section, we investigate the stability of numerical solutions to a non-interacting
gas of rigid dipoles.  Our strategy is based on convex optimization of the (1-point) density functional,
\begin{equation}
-\log Z[\phi] = \inf_F\sup_{\phi',\nu} \left[
    -n_0 (e^{\phi'_n} Z_L( m|\phi_e'| ) - 1) + \inner{\phi'-\phi}{F} + \nu (n - N) \right], \quad
Z_L(x) \equiv \sinh(x)/x
,\label{e:sad}
\end{equation}
in the representation of Eq.~\ref{e:fdip}.
The last term represents a constraint on the number density, with
$n$ understood to be a linear function of $F$ and $N$ a fixed constant.
Note that this problem has been phrased in a way that eliminates
taking the logarithm of the density, and makes it impossible for a converged
solution to violate the cone constraints.  Since $F = \pd{\log Z_1}{\phi}$,
the magnitude of the dipole density will always be smaller than its maximum,
$mn$.  These features contrast strongly with standard methods
based on using $F$ as the only control variable.

  Application of a simple primal-dual algorithm\cite{sboyd04} to this problem results
in a procedure identical to Newton-Rhapson iteration to find a zero of,
\begin{equation}
r(\phi', F, \nu) = [ F - \pd{\log Z_1(\phi')}{\phi'}, \phi' - \phi, n - N ]^T
.
\end{equation}
At each iteration, a backtracking search is made between
the starting point, $[\phi', F, \nu]$ ($t=0$) and the Newton-Rhapson
ending point ($t=1$, linear extrapolation to $r(\phi'+\delta\phi', F+\delta F, \nu+\delta\nu) = 0$).
The search scales down the step-size, $t$, by $1/2$ whenever $|r(t)| > (1 - 0.05 t) |r(0)|$.
When solving the linear system for the search direction, we cut the problem
size back to the dimensions of the original
by eliminating the equation for
$\delta \phi' = \text{diag}(\pd{^2\log Z_1}{\phi'^2})^{-1}( \delta F + r_{\phi'}(0))$ in,
\begin{equation}
\begin{bmatrix}
-\bar C dr + \text{diag}(\pd{^2\log Z_1}{\phi'^2})^{-1} & [1,0,0,0]^T \\
[1,0,0,0] dr & 0
\end{bmatrix}
\begin{bmatrix} \delta F \\ \delta \nu \end{bmatrix}
= - \begin{bmatrix} r_F(0) + \text{diag}(\pd{^2\log Z_1}{\phi'^2})^{-1} r_{\phi'}(0) \\ r_\nu(0) \end{bmatrix}.\label{e:NR}
\end{equation}
Although they are $1$ and $0$ in this example, this equation includes the integration
volume element, $dr$, and the direct correlation function, $\bar C$, to show the
general procedure for solving systems like Eq.~\ref{e:Dfunc}.

\begin{figure}
{\centering
\includegraphics[width=4in]{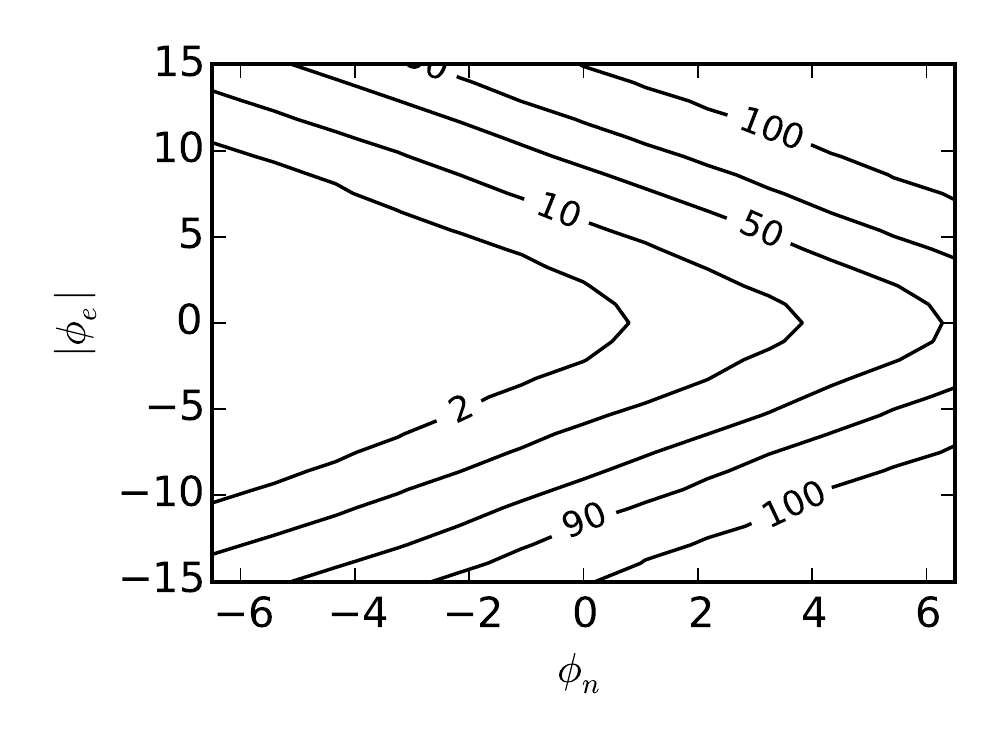}
\caption{Contour plot of the number of Newton-Rhapson
iterations required for convergence (to $10^{-8}$) of convex
optimization.
The increase in steps to convergence coincides with potentials
for which the equilibrium density is larger than 200.
Parameters used were $n_0 = 0.5, m = 1, \nu = 0$,
with $\phi_e$ along the 1,1,1 direction.}\label{f:steps}
}
\end{figure}

  Figure~\ref{f:steps} shows that the saddle-point iteration scheme
based on Eqns.~\ref{e:sad}-\ref{e:NR} is numerically robust for potentials
that result in small densities, $n$.  It achieves some of its
stability as a feature of Eq.~\ref{e:Z1}, which automatically
raises $n$ where there are strong electric fields.
Importantly, it shows breakdown is limited to large densities
where the scale of the problem is incongruent with
the convergence criteria (residual norm smaller than $10^{-8}$).
Even at $\phi = [7,8,8,8]^T$, where convergence was not obtained in 100 steps,
$n(r)$ increases linearly with iteration number.
One physical way to make the solution process more robust may be to truncate the summation
in Eq.~\ref{e:Z1} so as to place a grid-size-dependent
maximum on the local density.

  Of course, the 1-body problem can be easily solved analytically, but the goal
is to understand the numerical convergence behavior for molecular
fluids with more complicated terms and nonlocal interactions.
In those cases, the numerical issues remain the same, and
the only change is that the external potential, $\phi$, now includes
an additive, nonlocal linear term in $F$.
Our findings lead us to expect that convergence
in those cases will depend strongly on points in the fluid
where the potential is strongest (and the density is large).

\subsection{ Casimir Energy for Polarizable Point Dipoles}

  The inclusion of intramolecular degrees of freedom
allows a new approach to computing the dispersion free energy of solvating
a polarizable point dipole in a fluid.  This problem has
been raised in the ion solvation literature, where it has only previously
been possible to give a solution based on homogeneous dielectric theory.\cite{tduig13}

  We switch to a representation where $f = [1, p]^T$.  The coordinate $p$
is now a 3-dimensional point-dipole vector on each atom.  Its
default distribution is,
\begin{equation}
ds = (2\pi\sigma^2)^{-3/2} e^{-|p|^2/2\sigma^2} d^3p
,\label{e:gaus}
\end{equation}
associated with polarizability $\alpha = \beta \sigma^2$.
The integral $B = \int f f^\dagger ds = \text{diag}(1,\sigma^2,\sigma^2,\sigma^2)$
differs from the rigid rotator case, while the rest of the analysis in Sec.~\ref{s:wert}
remains unchanged. In particular, the definition of $h_\pm$ and $c_\pm$ should be the same as
Eqns.~\ref{e:hp} and~\ref{e:hm}, since the numerical factors
are associated with the Fourier transform over $r$.

  However, the appearance of $R$ in Eq.~\ref{e:MOZ} now requires,
\begin{align}
h_+ - c_+ &= 6K'\sigma^2 \rho_0 \, h_+ * c_+ \\
h_- - c_-  &= -3K'\sigma^2 \rho_0 \, h_- * c_-
,
\end{align}
where $\rho_0 = \avg{n}/V$, with the same boundary conditions on $h$ and $c$.

  Assuming the hypernetted chain closure ($\bar C = C$), the solvation
free energy functional contains isotropic and vector parts,
\begin{align}
-\log Z[\phi] &= \inf_{n} \Big\{ \mathcal F_n[\phi_I] + 
\inf_P \left[
\int dr \frac{|P(r)|^2}{2 n(r) \sigma^2}
-\tfrac{1}{2} \inner{P}{C_P*P}
- \inner{\phi_p}{P}
\right] \Big\} \label{e:FP} \\
\mathcal F_n[\phi_I] &\equiv \int dr \Big\{ n_0(r) - n(r) + n(r) \log\frac{n(r)}{n_0(r)} \Big\}
-\tfrac{1}{2} \inner{n - n_0}{c_I*(n-n_0)}
- \inner{\phi_I}{n}
.
\end{align}
The matrix $C_P$ is the lower-right $3\times 3$ sub-block of Eq.~\ref{e:Cdip}.
This form is particularly suitable for computational investigations
of implicit solvent.

  The $\mathcal F_n$ term describes normal excluded volume interactions
between the solute and solvent.  The other contribution to the solvation
free energy arises from electrostatic fluctuations, and describes a dispersion energy.
The inner minimization in Eq.~\ref{e:FP} has the solution,
\begin{equation}
- \tfrac{1}{2} \inner{\phi_P}{(I\delta(r, r')/\sigma^2 n(r) - C_P(r, r'))^{-1} * \phi_P}
.
\end{equation}

  The solvation free energy of a polarizable point dipole is the exponential
average of the solvation free energies of each of its possible configurational states
(Eq.~\ref{e:muex}).  We would like to use a dipole-dipole interaction energy,
$\phi_p(r) = \partial_r \, p\cdot \partial_r \frac{\beta}{4\pi\epsilon_0 |r|}$,
and integrate $Z[\phi_I, \phi_{p'}]$ over all possible solute polarizations, $p'$ using
the 1-molecule distribution from Eq.~\ref{e:gaus}.
However, $n(r)$ technically depends on the solute's polarization.  Ignoring this
dependence and using $Z$ instead of $Z_N$ (which ignores surface effects\cite{vball14})
yields an estimate of the solvation free energy for a polarizable dipole,
\begin{align}
-\log Z[\phi_I] &\approx \inf_{n} \Big\{
\mathcal F_n[\phi_I] + \tfrac{3}{2} \log \left(1 - q[n] \tfrac{\beta\sigma}{4\pi\epsilon_0}\right)
\Big\} , \\
\intertext{with}
q[n] &\equiv \int dr \int dr' \frac{r_x r'_x}{|r|^3 |r'|^3}
  \tr{ \partial_{r'}\otimes\partial_{r}  \left( I\delta(r, r')/\sigma^2 n(r) - C_P(r,r') \right)^{-1}(r,r') }
  .
\end{align}
This latter expression provides a new route to calculate the classical
dispersion energy of solvation.  Expanding the Green's function,
\begin{equation}
\left( I\delta(r, r')/\sigma^2 n(r) - C_P(r,r') \right)^{-1} = \sigma^2 I\delta(r,r') n(r') + \sigma^4 n(r) C_P(r,r') n(r') + O(\sigma^6)
,
\end{equation}
gives a Hamaker integral and a correlation correction as its first few terms.

\section{ Discussion}

  Molecular density functional theory has become a standard technique
in the statistical mechanics of fluids, but is not quite as popular as
as electronic density functional theory or the Poisson-Boltzmann
(or Debye-H\"{u}ckel theory) for electrolytes.
Part of this difficulty is due to a smaller body of literature
which does not focus on basis set expansions.
Our approach here follows a canonical route\cite{pdt2} by
developing the theory through the particle insertion expressions
(Eqns.~\ref{e:Z}-\ref{e:drho},\ref{e:muex}).  It contributes a careful discussion of the role
of basis functions as decomposing a resolution of identity, $\delta(x,x')$.
It also includes a sketch of free energy convergence issues that
are critical for numerical work,\cite{elomb95,ckell04,rishi13}
along with three diverse applications.

  Our presentation of the spherical Bessel translational basis expansions contrasts
with standard discussions of fast, grid-based spherical
Bessel transformations.\cite{jtalm09,mtoyo10}
Both provide a Fourier transform method for computing
convolutions of range-limited functions and require projection
of the result.  However, grid-based methods
use error analysis based on Riemann summation,
and leave some ambiguity on the proper treatment of the origin.\cite{jtalm09}
Using a space of analytical functions identifies three independent
sources of error which are understood in the sense of projections
-- approximating the original functions, and then projecting the
result of each multiplication and convolution.
It also has the advantage of strong results on numerical integrals.\cite{cfrap93}
Despite these advantages, the basis of spherical Bessel functions has the
same Gibbs' phenomenon issue as the Fourier transform -- making
it poorly approximate step functions.\cite{rcook28,ckell04}
Treating hard sphere solvation likely requires re-considering grid-based
methods, for example using a Haar basis of piecewise constant radial functions.

  The convolution algebra of 2-point correlation functions
developed in Sec.~\ref{s:conv} is a central idea of this work.
It enables compact expressions like the applications in Sec.~\ref{s:dip}, and hides the
complexities of new conformational information that can be
encoded in the basis (Sec.~\ref{s:basis}).
It can form the starting point for a simultaneous $N$-molecule optimization
working directly with $Q$ in Eq.~\ref{e:Q} in preference to
individual response functions, $F$.  This method could
then incorporate positive definite constraints on $Q$ required
for thermodynamic stability and existence of the direct correlation function, $C$.

\section{ Conclusions}

  It is clear that this approach can be usefully applied to new problems
in analytical and numerical solvation thermodynamics.
The concise derivations of Sec.~\ref{s:dip} were enabled by
translating the OZ equation to moment space.
Within this framework, we immediately
have Eq.~\ref{e:Dfunc} as an approximate expression for the density
functional.  Extending these results to alternative closures other than
the mean spherical approximation
is also simplified in the matrix formulation.

  The translation to moment space also simplifies analytical and
numerical investigations involving specialized forms of interaction energy.
Using Eq.~\ref{e:iQ2}, direct correlation functions for
higher-order multipolar interactions can be now computed straightforwardly
from correlations (Eq.~\ref{e:Q}) calculated in molecular dynamics simulations.
Several recent results (and computational implementations)\cite{droge18}
have tackled this inverse problem for dipolar solvents.\cite{gjean16}
The present work shows that further results of this type
can be made unambiguous by exactly specifying
which basis was used for $\mathcal H^\text{trs}$ and $f(q,s)$.
The complete theory above applies to any such basis.

  Section~\ref{s:basis} contains a novel presentation
of several possible basis expansions using the framework of
reproducing kernel Hilbert spaces.  This new perspective
treats inter- and intra-molecular
degrees of freedom in a holistic way.
It also provides analytical functional forms that can be
smoothly extrapolated to the origin (important for determining
macroscopic fluid properties) and are suitable
for investigating convergence of approximations.


  The formulation of the excess chemical potential using
Eq.~\ref{e:muex} goes beyond rigid molecule DFT by including
intramolecular degrees of freedom on the same footing as molecular
orientation.  This provides many new avenues for investigation on
the mathematical and applied theory of molecular solutions.


\begin{acknowledgements}
  We thank the USF Research Foundation,
NSF-XSEDE-CHE180030, and NSF-MRI-CHE-1531590
for support of this work.
\end{acknowledgements}

\bibliographystyle{spmpsci} 

\begin{thebibliography}{10}
\providecommand{\url}[1]{{#1}}
\providecommand{\urlprefix}{URL }
\expandafter\ifx\csname urlstyle\endcsname\relax
  \providecommand{\doi}[1]{DOI~\discretionary{}{}{}#1}\else
  \providecommand{\doi}{DOI~\discretionary{}{}{}\begingroup
  \urlstyle{rm}\Url}\fi

\bibitem{cahre09}
Ahrens, C., Beylkin, G.: Rotationally invariant quadratures for the sphere.
\newblock Proc. Royal Soc. A \textbf{465}(2110), 3103--3125 (2009).
\newblock \doi{10.1098/rspa.2009.0104}

\bibitem{aarch09}
Archer, A.J.: Dynamical density functional theory for molecular and colloidal
  fluids: A microscopic approach to fluid mechanics.
\newblock The Journal of Chemical Physics \textbf{130}(1), 014509 (2009).
\newblock \doi{10.1063/1.3054633}.
\newblock \urlprefix\url{https://doi.org/10.1063/1.3054633}

\bibitem{naron50}
Aronszajn, N.: Theory of reproducing kernels.
\newblock Trans. Amer. Math. Soc. \textbf{68}(3), 337--404 (1950).
\newblock \doi{10.1090/S0002-9947-1950-0051437-7}

\bibitem{vball14}
Ballenegger, V.: Communication: On the origin of the surface term in the
  {Ewald} formula.
\newblock J. Chem. Phys. \textbf{140}, 161102 (2014).
\newblock \doi{10.1063/1.4872019}

\bibitem{pdt2}
Beck, T.L., Paulaitis, M.E., Pratt, L.R.: Statistical thermodynamic
  necessities, chap.~2, pp. 23--31.
\newblock Cambridge, New York (2006)

\bibitem{dbegl97}
Beglov, D., Roux, B.: An integral equation to describe the solvation of polar
  molecules in liquid water.
\newblock J. Phys. Chem. B \textbf{101}(39), 7821--7826 (1997).
\newblock \doi{10.1021/jp971083h}

\bibitem{lblum90}
Blum, L., Cummings, P.T., Bratko, D.: A general solution of the molecular
  {Ornstein-Zernike} equation for spheres with anisotropic adhesion and
  electric multipoles.
\newblock J. Chem. Phys. \textbf{92}, 3741 (1990).
\newblock \doi{10.1063/1.457832}

\bibitem{lblum72}
Blum, L., Torruella, A.J.: Invariant expansion for two-body correlations:
  Thermodynamic functions, scattering, and the {Ornstein-Zernike} equation.
\newblock J. Chem. Phys. \textbf{56}(303), 303--310 (1972).
\newblock \doi{10.1063/1.1676864}

\bibitem{sboch54}
Bochner, S.: Positive zonal functions on spheres.
\newblock Proc. Nat. Acad. Sci. USA \textbf{40}, 1141--1147 (1954)

\bibitem{sboyd04}
Boyd, S., Vandenberghe, L.: Convex Optimization.
\newblock Cambridge Univ. Press, Cambridge UK (2004)

\bibitem{pbuen05}
Buenzli, P.R., Martin, P.A.: Microscopic origin of universality in {Casimir}
  forces.
\newblock J. Stat. Phys. \textbf{119}(1--2), 273--307 (2005).
\newblock \doi{10.1007/s10955-004-1990-4}

\bibitem{rcook28}
Cooke, R.G.: {Gibbs's} phenomenon in {Fourier-Bessel} series and integrals.
\newblock Proc. London Math. Soc. pp. 171--192 (1928).
\newblock \doi{10.1112/plms/s2-27.1.171}

\bibitem{davis}
Davis, H.T.: Statistical Mechanics of Phases.
\newblock VCH Publishers, New York (1996)

\bibitem{hdett97}
Dette, H., Studden, W.J.: The Theory of Canonical Moments with Applications in
  Statistics, Probability, and Analysis.
\newblock Wiley, New York (1997)

\bibitem{lding17}
Ding, L., Levesque, M., Borgis, D., Belloni, L.: Efficient molecular density
  functional theory using generalized spherical harmonics expansions.
\newblock J. Chem. Phys. \textbf{147}, 094107 (2017).
\newblock \doi{10.1063/1.4994281}

\bibitem{tduig13}
Duignan, T.T., Parsons, D.F., Ninham, B.W.: A continuum solvent model of the
  multipolar dispersion solvation energy.
\newblock J. Phys. Chem. B \textbf{117}(32), 9412--9420 (2013).
\newblock \doi{10.1021/jp403595x}

\bibitem{kdyer08}
Dyer, K.M., Perkyns, J.S., Stell, G., Pettitt, B.M.: A molecular site-site
  integral equation that yields the dielectric constant.
\newblock J. Chem. Phys. \textbf{129}(10), 104512 (2008).
\newblock \doi{10.1063/1.2976580}

\bibitem{uessm95}
Essmann, U., Perera, L., Berkowitz, M.L., Darden, T., Lee, H., Pedersen, L.G.:
  A smooth particle mesh {Ewald} method.
\newblock J. Chem. Phys. \textbf{103}, 8577--8592 (1995)

\bibitem{cfrap93}
Frappier, C., Olivier, P.: A quadrature formula involving zeros of {Bessel}
  functions.
\newblock Math. Comput. \textbf{60}(201), 303--31 (1993)

\bibitem{pfrie86}
Fries, P.H., Patey, G.N.: The solution of the {Percus-Yevick} approximation for
  fluids with angle-dependent pair interactions. a general method with results
  for dipolar hard spheres.
\newblock J. Chem. Phys. \textbf{85}, 7307 (1986).
\newblock \doi{10.1063/1.451369}

\bibitem{eshrig}
von Helmut~Eschrig: Legendre transformation.
\newblock In: The Fundamentals of Density Functional Theory,
  \emph{Teubner-Texte zur Physik}, vol.~32, pp. 99--126. B. G. Teubner
  Verlagsgesellschaft, Leipzig (1996)

\bibitem{mikeg95}
Ikeguchi, M., Doi, J.: Direct numerical solution of the {Ornstein-Zernike}
  integral equation and spatial distribution of water around hydrophobic
  molecules.
\newblock J. Chem. Phys. \textbf{103}(12), 5011--5017 (1995).
\newblock \doi{10.1063/1.470587}

\bibitem{rishi13}
Ishizuka, R., Yoshida, N.: Extended molecular {Ornstein-Zernike} integral
  equation for fully anisotropic solute molecules: Formulation in a rectangular
  coordinate system.
\newblock J. Chem. Phys. \textbf{139}(8), 084119 (2013).
\newblock \doi{10.1063/1.4819211}

\bibitem{gjean16}
Jeanmairet, G., Levy, N., Levesque, M., Borgis, D.: Molecular density
  functional theory of water including density-polarization coupling.
\newblock J. Phys.: Condens. Matter \textbf{28}, 244005 (2016).
\newblock \doi{10.1088/0953-8984/28/24/244005}

\bibitem{jjohn16}
Johnson, J., Case, D.A., Yamazaki, T., Gusarov, S., Kovalenko, A., Luchko, T.:
  Small molecule hydration energy and entropy from {3D-RISM}.
\newblock J. Phys.: Condens. Matter \textbf{28}, 344002 (2016).
\newblock \doi{10.1088/0953-8984/28/34/344002}

\bibitem{ckell04}
Kelley, C.T., Pettitt, B.M.: A fast solver for the {Ornstein-Zernike}
  equations.
\newblock J. Comput. Phys. \textbf{197}(2), 491--501 (2004).
\newblock \doi{10.1016/j.jcp.2003.12.006}

\bibitem{skony98}
Konyaev, S.I.: Quadrature formulae invariant to the symmetry group of a
  24-hedron for a spherically symmetric weight function in $\mathbb r^4$.
\newblock Russian J. Numer. Analysis and Math. Model. \textbf{13}(1), 13--26
  (1998).
\newblock \doi{10.1515/rnam.1998.13.1.13}

\bibitem{akova00}
Kovalenko, A., Hirata, F.: Potentials of mean force of simple ions in ambient
  aqueous solution. i. three-dimensional reference interaction site model
  approach.
\newblock J. Chem. Phys. \textbf{112}, 10391 (2000).
\newblock \doi{10.1063/1.481676}

\bibitem{elomb95}
Lomba, E., L\'{o}pez-Mart\'{\i}n, J.L.: On the solutions of the hypernetted
  chain equation inside the gas-liquid coexistence region.
\newblock J. Stat. Phys. \textbf{80}(3--4), 825--839 (1995).
\newblock \doi{0022-4715/95/0800-0825}

\bibitem{wmaje14}
Majewski, W.A., Labuschagne, L.E.: On applications of {Orlicz} spaces to
  statistical physics.
\newblock Ann. Henri Poincar\'{e} \textbf{15}, 1197--1221 (2014).
\newblock \doi{10.1007/s00023-013-0267-3}

\bibitem{nrft}
Matyushov, D.V.: Solvent reorganization energy of electron-transfer reactions
  in polar solvents.
\newblock J. Chem. Phys. \textbf{120}, 7532 (2004).
\newblock \doi{10.1063/1.1676122}

\bibitem{dmaty94}
Matyushov, D.V., Ladanyi, B.M.: A perturbation theory and simulations of the
  dipole solvation thermodynamics: Dipolar hard spheres.
\newblock J. Chem. Phys. \textbf{110}, 994 (1999).
\newblock \doi{10.1063/1.478144}

\bibitem{gmaze06}
Mazenko, G.F.: Nonequilibrium Statistical Mechanics.
\newblock Wiley-VCH (2006)

\bibitem{frain01}
Raineri, F.O., Stell, G.: Dielectrically nontrivial closures for the {RISM}
  integral equation.
\newblock J. Phys. Chem. B \textbf{105}(47), 11880--11892 (2001).
\newblock \doi{10.1021/jp0121163}

\bibitem{rrami02}
Ramirez, R., Gebauer, R., Mareschal, M., Borgis, D.: Density functional theory
  of solvation in a polar solvent: Extracting the functional from homogeneous
  solvent simulations.
\newblock Phys. Rev. E \textbf{66}, 031206 (2002).
\newblock \doi{10.1103/PhysRevE.66.031206}

\bibitem{rao}
Rao, M.M., Ren, Z.D.: Theory of {Orlicz} spaces, \emph{Monographs and textbooks
  in pure and applied mathematics}, vol. 146.
\newblock M. Dekker, New York (1991)

\bibitem{droge15}
Rogers, D.M.: Real-space quadrature: a convenient, efficient representation for
  multipole expansions.
\newblock J. Chem. Phys. \textbf{142}, 074101 (2015)

\bibitem{droge18}
Rogers, D.M.: Extension of {Kirkwood-Buff} theory to the canonical ensemble.
\newblock J. Chem. Phys. \textbf{148}, 054102 (2018).
\newblock \doi{10.1063/1.5011696}

\bibitem{svolo12}
Sergiievskyi, V.P., Fedorov, M.V.: 3drism multigrid algorithm for fast
  solvation free energy calculations.
\newblock Journal of Chemical Theory and Computation \textbf{8}(6), 2062--2070
  (2012).
\newblock \doi{10.1021/ct200815v}.
\newblock \urlprefix\url{https://doi.org/10.1021/ct200815v}.
\newblock PMID: 26593838

\bibitem{tsluc81}
Sluckin, T.J.: Density functional theory for simple molecular fluids.
\newblock Mol. Phys. \textbf{43}(4), 817--849 (1981).
\newblock \doi{10.1080/00268978100101711}

\bibitem{jtalm09}
Talman, J.D.: {NumSBT}: A subroutine for calculating spherical {Bessel}
  transforms numerically.
\newblock Comput. Phys. Commun. \textbf{180}(2), 332--338 (2009).
\newblock \doi{10.1016/j.cpc.2008.10.003}

\bibitem{mtoyo10}
Toyoda, M., Ozaki, T.: Fast spherical {Bessel} transform via fast {Fourier}
  transform and recurrence formula.
\newblock Comput. Phys. Commun. \textbf{181}, 277--282 (2010).
\newblock \doi{10.1016/j.cpc.2009.09.020}

\bibitem{lvrbk09}
Vrbka, L., Lund, M., Kalcher, I., Dzubiella, J., Netz, R.R., Kunz, W.:
  Ion-specific thermodynamics of multicomponent electrolytes: A hybrid {HNC/MD}
  approach.
\newblock J. Chem. Phys. \textbf{131}, 154109 (2009)

\bibitem{mwert71}
Wertheim, M.S.: Exact solution of the mean spherical model for fluids of hard
  spheres with permanent electric dipole moments.
\newblock J. Chem. Phys. \textbf{55}, 4291 (1971).
\newblock \doi{10.1063/1.1676751}

\bibitem{bwido82}
Widom, B.: {Potential-distribution theory and the statistical mechanics of
  fluids}.
\newblock J. Phys. Chem. \textbf{86}, 869--872 (1982)

\bibitem{szhao13}
Zhao, S., Liu, H., Ramirez, R., Borgis, D.: Accurate evaluation of the
  angular-dependent direct correlation function of water.
\newblock J. Chem. Phys. \textbf{139}, 034503 (2013)

\end{thebibliography}

\end{document}